\documentclass[11pt]{amsart}

\usepackage[utf8]{inputenc}
\RequirePackage{fix-cm}
\usepackage{graphicx}
\usepackage{svg}
\usepackage{amsmath}
\usepackage{amssymb}
\usepackage{float}
\usepackage{array,colortbl,multirow,multicol,booktabs,ctable}
\usepackage{chngpage}

\usepackage{graphicx}
\DeclareMathOperator{\Workflow}{\mathbf{Workflow}}
\DeclareMathOperator{\Value}{\mathbf{Value}}
\DeclareMathOperator{\Information}{\mathbf{Information}}
\DeclareMathOperator{\Quality}{\mathbf{Quality}}
\DeclareMathOperator{\Actionability}{\mathbf{Actionability}}
\DeclareMathOperator{\Accuracy}{\mathbf{Accuracy}}
\DeclareMathOperator{\Sources}{\boldsymbol{\#}\mathbf{Sources}}
\DeclareMathOperator{\Cost}{\mathbf{Cost}}
\DeclareMathOperator{\Complexity}{\mathbf{Complexity}}
\DeclareMathOperator{\dwelltime}{\mathbf{dwell}\text{ }\mathbf{time}}
\DeclareMathOperator{\frameintegration}{\mathbf{frame}\text{ }\mathbf{integration}}

\DeclareMathOperator*{\argmin}{\arg\!\min}
\DeclareMathOperator*{\argmax}{\arg\!\max}

\begin{document}


\title{A Framework for the Optimal Selection for High-Throughput Data Collection Workflows by Autonomous Experimentation Systems}


\author{Rohan Casukhela}
\address{$^1$Department of Materials Science \& Engineering, The Ohio State University,
              2041 College Rd N, Columbus, OH 43210, USA}
\email{rcasukhela@outlook.com}

\author{Sriram Vijayan}
\address{$^1$Department of Materials Science \& Engineering, The Ohio State University,
              2041 College Rd N, Columbus, OH 43210, USA}
\email{vijayan.13@osu.edu}

\author{Joerg R. Jinschek}
\address{$^1$Department of Materials Science \& Engineering, The Ohio State University,
              2041 College Rd N, Columbus, OH 43210, USA}
\address{$^2$DTU Nanolab, Technical University of Denmark, Kgs. Lyngby, DK}
\email{jojin@dtu.dk}

\author{Stephen R. Niezgoda}
\address{$^1$Department of Materials Science \& Engineering, The Ohio State University,
              2041 College Rd N, Columbus, OH 43210, USA}
\address{$^3$Department of Mechanical \& Aerospace Engineering, 201 W 19th Ave, Columbus, OH 43210, USA}
\email{niezgoda.6@osu.edu}

\begin{abstract}
 Autonomous experimentation systems have been used to greatly advance the integrated computational materials engineering (ICME) paradigm. This paper outlines a framework that enables the design and selection of data collection workflows for autonomous experimentation systems. The framework first searches for data collection workflows that generate high-quality information and then selects the workflow that generates the \emph{best, highest-value} information as per a user-defined objective. We employ this framework to select the \emph{user-defined best} high-throughput workflow for material characterization on an additively manufactured Ti-6Al-4V sample for the purposes of outlining a basic materials characterization scenario, reducing the collection time of backscattered electron scanning electron scanning electron microscopy images by a factor of 5 times as compared to the benchmark workflow for the case study presented, and by a factor of 85 times as compared to the workflow used in the previously published study.
\end{abstract}



\maketitle




\section{Introduction}
\label{intro}
Autonomous experimentation (AE) (including autonomous simulation) is being explored as a strategy to accelerate materials design and reduce product development cycles [\cite{stach2021autonomous}, \cite{flores2020materials}, \cite{szymanski2021toward}, \cite{roch2018chemos}, \cite{montoya2022toward}]. Autonomous experimentation is defined by Stach et al. as “…an iterative research loop of planning, experiment, and analysis…” [\cite{stach2021autonomous}. Materials AE research is a rapidly advancing field. Powerful applications of AE in materials science include implementing Bayesian optimization principles in AE systems to quickly optimize material properties of interest [\cite{ling2017high}, \cite{lookman2017statistical}, \cite{balachandran2020adaptive}, \cite{xu2016heterogeneous}, \cite{christensen2021data}, \cite{hase2021olympus}], and creating AE systems to perform high-throughput experimentation (HTE) for rapid materials discovery and optimization for polymers, metals, ceramics, and more [\cite{ong2019accelerating}, \cite{calderon2015aflow}, \cite{maier2019early}, \cite{vasudevan2019materials}, \cite{curtarolo2012aflowlib}, \cite{hattrick2016perspective}, \cite{liu2019high}, \cite{pollice2021data}].

We consider AE in a broader context and pose a futuristic scenario where scientific discovery proceeds from a human investigator giving a simple command to a computer, such as identifying the likely root cause of failure for an example component. While this thought experiment borders on science fiction, it is instructive to consider the steps the autonomous system must complete to arrive at a final conclusion. For this autonomous exploration to be carried out, the system must: 

\begin{enumerate}
    \item Parse the verbal instructions into a quantifiable objective that meets the requirement of the user
    \item Identify information required to achieve the objective
    \item Design a workflow collect relevant information. In the materials realm this might include, for example, a sequence of testing, characterization, and simulation steps
    \item Design a sequence of experiments using the workflow from Step 3 to optimally gain information about the system
    \item Execute experiments to collect data
    \item Extract information from data and assess if objective is met
    \item Iterate on Steps 3-6 if objective is not met
\end{enumerate}

Steps 1 and 2 fall in the realm of knowledge discovery via natural language processing (NLP) [\cite{choudhary2022recent}, \cite{venugopal2021looking}, \cite{shetty2021automated}, \cite{dieb2021creating}]. Steps 4 through 7 fall in the realm of optimal experiment design [\cite{ling2017high}, \cite{lookman2017statistical}, \cite{balachandran2020adaptive}, \cite{xu2016heterogeneous}, \cite{christensen2021data}, \cite{hase2021olympus}]. Both tasks are active areas in research. Step 3 involves the design of workflows. The focus of this writing is on expanding the capabilities of modern AE systems to complete Step 3 independent of human guidance or intervention.  Here, we define a workflow as: \emph{``the set of procedures, methods, and models used to observe physical/virtual systems''}. The working assumption here is that a specific objective set in Step 1 (e.g. grain size measurement, tensile stress/strain curve….) is established and potential experimental and simulation-based tools identified in Step 2 (e.g. microscopy imaging, hardness testing devices, heat flow models, FEM simulations, etc.) are in place. Workflow design consists of determining how best to use these experimental and simulation tools to collect relevant data to the objective. From here, Steps 4 through 7 can proceed normally, returning to Step 3 when necessary.

 During materials/process development cycles, Step 3 is challenging for AE systems as it requires human-like domain knowledge of engineering and materials systems, as well as engineering properties of interest. Once the type of information to be gathered is ascertained, the natural next question is “How do we actually collect that information?” Current AE efforts start with the adoption of a human-designed workflow that remains static throughout the entire process. While some of these workflow decisions may seem trivial to human experts, even the most basic experimental procedures are often outlined by detailed standards and operating procedures. Often, investigators will simplify a complex procedure with the understanding that the quality of the results does not significantly change. These modifications to procedures are often made to maximize repeatability, reduce time or cost of data acquisition, or account for different sources of variability. For example, unless testing is being performed for certification and qualification, most tensile testing is not performed to ASTM E8 specifications, even though the output of these tests is largely acceptable and reliable--tensile testing of miniature specimens is one such occurrence. Another example is materials characterization by scanning electron microscope (SEM): a human scientist operating an SEM relies on their prior experiences and intuition to determine the method of sample preparation, magnification, field of view, accelerating voltage, beam current, contrast/brightness, and so on. It is a non-trivial question for the AE system to assess when a given procedure is “good enough” for the scientific objectives given resource constraints, even when the data objectives and basic procedures have already been delineated in Step 2 above.

 Current advances in materials AE involve \emph{a priori} selecting one workflow to measure the system, despite the possible existence of other workflows that might yield a higher-certainty, higher-accuracy measurement at lower cost of acquisition. An analogue to this issue can be seen in autonomous vehicles. An autonomous vehicle uses sensor networks to make decisions without human intervention. However, autonomous vehicles do not decide what sensors to use or ignore in a given scenario. The priority and properties of the signals are predefined by the engineer. Human scientists and engineers design both the sensor networks and the vehicle, thus dictating how the vehicle gathers and uses information. This approach has three notable advantages:

\begin{enumerate}
    \item The data stream from the setup is controlled and predictable
    \item The AE system can quickly iterate through the objective space without having to potentially change tooling or account for the difference in measurements to the objective space as the workflow is changed
    \item The measurement process is repeatable and thus makes AE easily scale to required levels
\end{enumerate}

 The notable disadvantages of using a static workflow are that changes to the workflow or potential improvements cannot be quantified by the AE system, and that a human must define the best method for information collection rather than the AE system. Current approaches to the design of materials AE systems severely limit the potential application space of materials AE systems to tasks that are strictly defined by human operators, repetitive in nature, and limited in scope. Due to the growing activity in materials AE, AE systems need to be designed so that they may quickly adapt to rapid advances made in experimental/simulation/data-processing technologies [\cite{choudhary2022recent}]. Hence, advancing the decision authority of AE systems is crucial for design of next-generation materials AE systems and maintaining the relevancy of already existing materials AE systems.

 In order to enable AE systems to discover the \emph{user-defined best} (high-value) data collection workflows independent of human scientists and engineers, we pose a framework reminiscent of multi-objective optimization techniques to dynamically select the highest-value workflow that generates structured materials data:

\begin{enumerate}
    \item An objective is established by the user to guide workflow development
    \item The procedures, methods, and models are listed that will be considered in the workflow
    \item A fast search over the space of possible workflows is conducted to quickly target high-quality workflows in the context of the objective
    \item A high-resolution search over high-quality workflows is conducted to select a high-value workflow
\end{enumerate}

 The concept of this framework will be described in detail, then illustrated in a case study. There, the impact of a deep-learning based denoising algorithm on a materials characterization workflow is examined. The framework was used to algorithmically select the \emph{user-defined best} high-throughput workflow that collects backscattered electron scanning electron microscope (BSE-SEM) images on the material sample at approximately 85 times greater speed compared to the previous study [\cite{shao2020effect}], and 5 times faster than the ground-truth workflow of the presented case study. Lastly, summary statistics for the data stream of the selected workflow are provided.

\section{Workflow Design Framework}
\label{Framework}
\subsection{Motivation}
\label{Motiv}
 All data collection efforts must begin with specifying an objective that needs to be met. A well-designed $\Workflow$ maximizes objective-dependent $\Value$ by providing insightful $\Information$ to assist in making a decision. $\Information$ can then be extracted from the collected data, which ultimately provides $\Value$ according to the objective.

\begin{equation}
\Workflow \rightarrow \Information \rightarrow \Value
\end{equation}

 We use the direct product of the workflow, the extracted information, as a measure of the value of the workflow itself. Extracted information is a direct summary of the raw data (number of pores within an image, average number of hairline cracks in a part, etc.), that describes the system under investigation. If the extracted information is high-accuracy and high-certainty, the use of black-box data processing/transformation methods is justified in any workflow that utilizes them. Thus, the subtleties of how information is extracted from data can safely be ignored, which naturally allows for the use of less interpretable classes of models (e.g. neural networks) in the workflow.
 
 The $\Value$ of information is proportional to the information’s $\Quality$ and $\Actionability$.
\begin{equation}
\Value \propto \Quality \cap \Actionability
\end{equation}

 \noindent $\Actionability$ tells us how useful information is in achieving a particular objective. High-actionability information is critical to making high-value decisions. The ground-truth defect density of a part to estimate the overall mechanical stability in a critical application is information of high-actionability. In some cases, collecting highly-actionable information, such as the ground-truth, can be expensive or intractable. In these cases, the cost for collecting this information must be assumed to be infinite and other, lower-actionable types of information must be sought. In contrast to high-actionability information, low-actionability information is less useful to making high-value decisions.

 The $\Quality$ of information is proportional to its $\Accuracy$ with respect to a pre-determined ground truth and the number of unique data sources, $\Sources$, from which it is harvested, while being inversely proportional to the $\Cost$ of acquisition.

\begin{equation}
\Quality \propto \frac{\Accuracy \cap \Sources}{\Cost}
\end{equation}

In general, increasing the $\Accuracy$ and/or the $\Sources$ reduces the uncertainty about the system that is under investigation. Therefore, the two quantities are related to the amount of valuable information that the workflow generates. However, increased $\Accuracy$ and the $\Sources$ typically leads to increased $\Cost$ of acquisition. This is due to the extra time needed and the extra resources required to collect, structure and curate the data [\cite{casukhela2021towards}].

 The approach for any effort should be to select the highest-value workflow from a set of high-quality data collection workflows. High-quality workflows strike a balance between generating information that is accurate or high-certainty, and generating information that is low-cost. A high-quality workflow that generates high-actionability information is considered to be a high-value workflow. As an example, high-throughput experimental workflow design aims to select a workflow that is considered high-quality and generates a given amount of information in the shortest amount of time at the lowest possible cost. In this study, we show that the setup and design of workflows can be addressed by treating them as a two-stage optimization problem.

\subsection{Mathematical Definition of Framework}
\label{def}
 It is very challenging to find the highest-value experimental setup in one step. Therefore, we conduct the search for the highest-value workflow in a two-stage approach: In \emph{Stage I} we seek to identify the set of workflows that maximize the quality of the information. In \emph{Stage II} we select the highest-value workflow from the set of workflows obtained in \emph{Stage I} by maximizing actionability (see Fig 1).

\begin{figure}[H]
\centering
\includegraphics[scale=0.45]{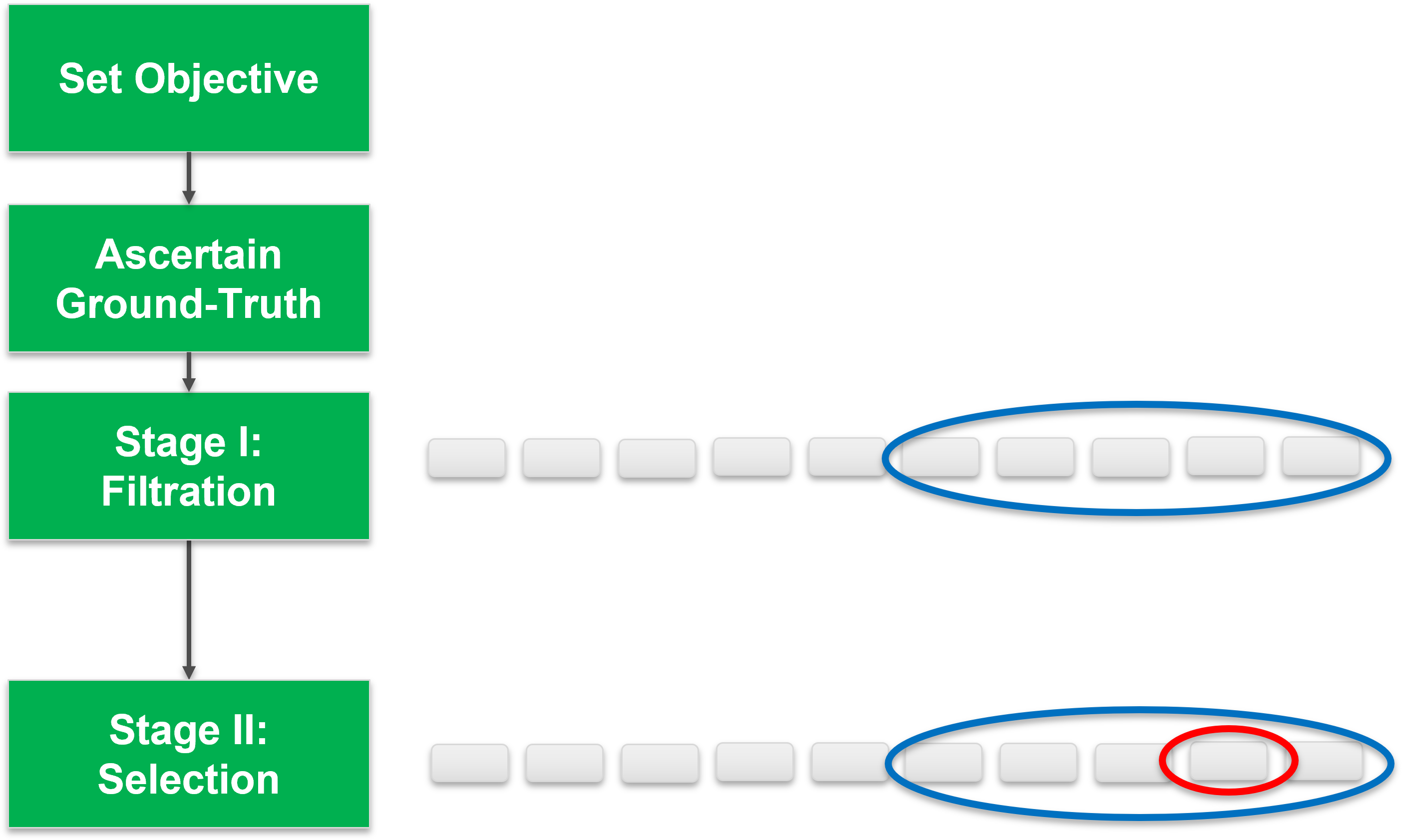}
\caption{Framework schematic showing the two-stage approach when searching for the highest-value workflow.}
\label{fig:schem}
\end{figure}

Let $x$ be the data obtained from the collection process, dependent on the process collection settings $\theta$, $A$ be the data processing sequence that is applied to data $x(\theta)$, with data processing parameters $\lambda$, the product $A_{\lambda} x(\theta)$ be the extracted information from the workflow, and $M$ be the design specification, or design parameter that we designate as ground-truth.

\begin{multline}
\text{\emph{Stage I}}: \quad \theta^*, \lambda^* = \argmin_{\theta, \lambda}
\Big( C_1 \vert\vert (A_\lambda x(\theta) - M\vert\vert + C_2 \Cost(x(\theta))\\
 + C_3 \Complexity (A_\lambda x(\theta))\Big),\quad \lbrace C_1, C_2, C_3 > 0\rbrace
\end{multline}

\begin{multline}
\text{\emph{Stage II}}: \argmax_{\theta^*, \lambda^*} \big( \Actionability (x(\theta), A_\lambda x(\theta) \Big)\\
\text{report bias, and standard deviation of setup.}
\end{multline}

\begin{itemize}
    \item $A_\lambda x(\theta)$ and $M$ are compared in the \emph{Stage I} objective function.
	\item $\Cost$ is a term that accounts for how expensive it is to collect the data $x$.
	\item $\Complexity$ is a term that accounts for the workflow A’s complexity, computation time/resources, curation time (which can increase due to the number of sources), and interpretability. A simple example would be to use the term $p + q^2$, where $p$ approximates the time/space complexity of algorithms used, and $q$ corresponds to the number of sources of data.
	\item $C_1, C_2,$ and $C_3$ \emph{are user-defined}, and are weights that can be used to adjust the importance of each term in the objective function. Larger numbers imply more importance, and smaller numbers imply less importance.
\end{itemize}

 In \emph{Stage I} we find a set of candidate experimental setups with process collection settings process collection settings $\theta^*$ and data processing parameters $\lambda^*$ that minimize the objective function. It guarantees the quality of the workflow in the first place, ensuring that only workflows that generate high-quality information remain. For high-throughput workflow design, $\Actionability$ can be defined as collecting the most amount of data in a given amount of time. We use this definition of $\Actionability$ for our \emph{Stage II} criteria in our case study (see Sect.~\ref{case}).

 \emph{Stage II} directly compares the candidate setups with process collection settings process collection settings, $\theta^*$, and data processing parameters, $\lambda^*$, and finds the setup among these that maximizes the $\Actionability$ of the workflow.

 The \emph{Stage I} objective function describes the quality of the information generated by the workflow: smaller scores of the function will correspond to a higher-quality workflow, while larger scores of the function will correspond to a lower-quality workflow. The goal, of course, is to minimize the objective function as much as possible. Regions of the same objective values, outline \emph{iso-quality} regions. Iso-quality regions imply that both workflow settings produce information of the same quality. For the design of high-throughput workflows, we select the setting that yields more data (i.e. minimize acquisition time), between two settings that have approximately the same, minimal \emph{Stage I} objective function value.

 The \emph{Stage I} objective function is inherently “noisy”, meaning that repeated measurement of the experimental setups will output different \emph{Stage I} objective values. This is the reason why \emph{Stage II} is required: \emph{Stage I} is a coarse search that informs the user of what regions to investigate further. \emph{Stage II} is a fine search and guides the user to quantify the accuracy and variance of the different setups. \emph{Stage II} allows for the possibility that \emph{Stage I’s} objective function values are random. In practice, \emph{Stage II} takes candidate setups with similar objective function values as input and compares the setups. If a meaningful difference between the setups is found, a decision can be made to select the setup closest in mean to the ground-truth setup that also minimizes acquisition time.

 To use this framework, experiments should first be conducted to seek a set of well-established, well-understood “ground-truth” measurements that can be used to compare the workflow’s ability to produce accurate measurements. Second, the accuracy, cost, and analysis complexity term weights $C_1,C_2,C_3$ should be properly defined. Third, a \emph{Stage I} search should be conducted. This can be done by collecting extracted information across all workflows being considered and using the extracted information to score each workflow. It is recommended that very few measurements (one measurement per setup is permissible) for each setup are collected in this step, as the \emph{Stage I} exploration space can be large. \emph{Stage I} will filter the workflow space and produce a set of candidate workflows. Workflows that minimize the objective function or are close to the minimum should be included within the candidate set. Fourth, a \emph{Stage II} data collection effort using the candidate workflows for finer comparison should be conducted. Because a \emph{Stage II} search needs to be higher-resolution than \emph{Stage I}, it will require more data per workflow than in the \emph{Stage I} search. The \emph{Stage II} search can be conducted via a formal statistical experimental design, such as one-way ANOVA. The \emph{Stage II} experiment should be designed so that it tests for any difference between the proposed experimental setups and the ground-truth setups. This fourth and final step will yield the highest-value data collection workflow.

\section{Case Study: Designing a High-Throughput Workflow for Expediting Microstructural Characterization of AM Builds}
\label{case}

 We now present a case study in which we design a high-throughput workflow for the microstructural characterization of an additively manufactured (AM) sample using BSE-SEM images. The AM sample was manufactured in a previous study that examined and compared the structure-property relationships between parts that had slightly different beam scanning strategies. The part used for this study was fabricated using a Linear Scan (LS) strategy, and the sample was sectioned from the top-center area of the part (see details in [\cite{shao2020effect}]).

 Using SEM characterization is particularly interesting in the research of high-throughput workflows for materials applications because SEMs can be used to investigate a materials’ microstructure and generate large amounts of high-value data sets. Thus, developing high-throughput workflows involving SEMs is a highly desirable goal for material structure-property prediction.

 An SEM image or frame is acquired in a pixel-by-pixel approach, while the number of pixels has an upper limit determined by the SEM’s scan engine. The electron beam rasters in lines over an area of interest step-by-step (pixel-by-pixel) with an operator-chosen rest time (i.e. dwell time) at each of the steps. This rest time has a huge impact on the quality (here: signal-to-noise, SNR) of the final image. An operator-chosen image magnification determines the number and size of the pixels in relation to the feature size in the sample (e.g., size of a grain or a defect) that is of interest. In principle, the image acquisition time is determined by the chosen number of pixels in an image multiplied by the chosen dwell time per pixel. Integration of multiple fast-scanned images (i.e. frame integration, FI) can be used to suppress noise. The constraints are that the pixel size needs to be optimal in relation to the size of a feature that needs to be resolved, and the dwell time needs to be sufficiently high to achieve a required SNR to recognize the feature in the subsequent data analysis process.

 The trade-off between dwell time and frame integration (FI) vs. image quality is an example of a compromise one has to accept when using a high-throughput approach (see Fig. 2): decreasing the cost of data acquisition (i.e. taking images faster with a shorter dwell time) decreases image quality and might therefore lead to greater deviations from a specified $M$.

\begin{figure}[H]
\centering
\includegraphics[scale=0.65]{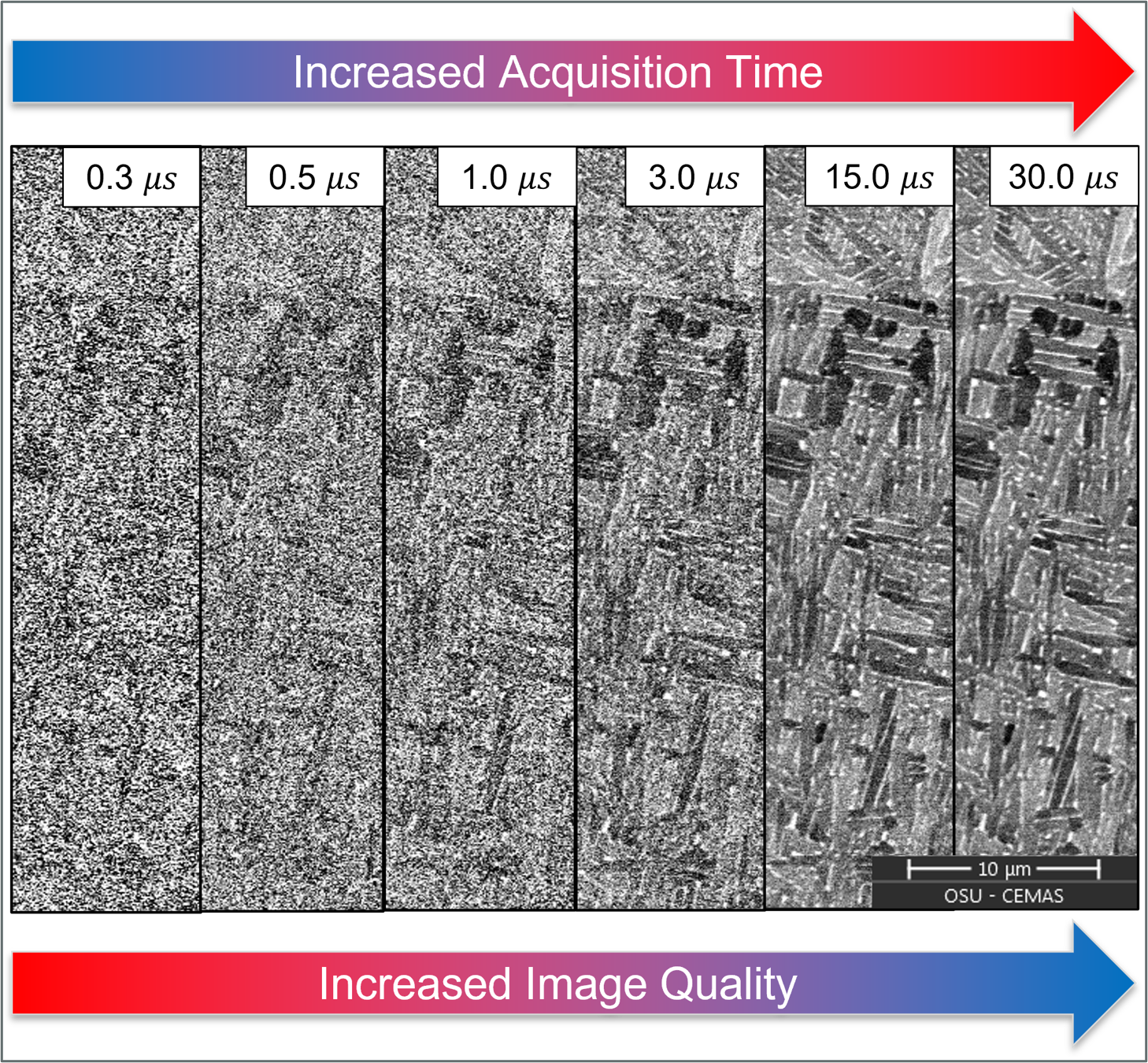}
\caption{Backscattered electron (BSE) scanning electron microscopy (SEM) images taken with increasing dwell time (1 FI to 32 FI indicates the number of frames used in frame integration) visualizing one of the many trade-offs between acquisition time vs. image quality.}
\label{fig:tradeoff}
\end{figure}

 Recent advances have shown that deep-learning algorithms can be used to boost the SNR of low-quality microscopy images and yield images with effectively higher SNR [\cite{ronneberger2015u}, \cite{mohan2020deep}, \cite{de2019resolution}, \cite{roels2020interactive}, \cite{zhang2020denoising}, \cite{zeltmann2020denoising}, \cite{mohan2022deep}]. These methods have exhibited robust performance on images outside of the testing dataset when compared to more standard denoising techniques. Deep learning-based denoising algorithms provide a straightforward solution to the aforementioned problem of taking SEM images in a high-throughput fashion. Now, one can take images with low SNR (or fewer pixels) and still retain the quality of higher-acquisition-time, higher-signal, higher-resolution images. The question then becomes, "by how much does our process improve?"

 In this case study, sample and data used was the linear scan XY top Ti-6Al-4V (Ti64) metal alloy sample, previously reported in Shao et al [\cite{shao2020effect}]. The objective is to design and implement a high-throughput workflow utilizing an algorithm that denoises BSE-SEM images to systematically characterize an AM part. The extracted information for this case study will be the size of the Ti64 $\alpha$-lath (i.e. $\alpha$-lath thickness).

\subsection{Material Sample}
\label{sample}
Details about sample fabrication have previously been reported [\cite{shao2020effect}]. The center region of the linear scan top XY sample was used for this case study (see figure 3a in Shao et al).

\subsection{Denoising Algorithm}
\label{denoise}
The BSE denoiser model presented here is a convolutional neural network (CNN) with the U-Net architecture. Further model details can be found at the following references [\cite{ronneberger2015u}, \cite{thermoxtra}].

\subsection{Stage I Search}
\label{case:stage-i}
 Here, $C_1$ is 500, the distance term between $Ax$ and $M$ is defined as the squared distance between the two terms, the cost term is defined by a combination of terms that relate to the acquisition time of a micrograph, and the model complexity term is set to zero. The model complexity term is set to zero because all images for this case study had the same pixel resolution, and therefore the resources needed to process all images was the same. Therefore, the model complexity term is constant and does not affect the minimization of our \emph{Stage I} objective function for this case study.

 The objective function:

\begin{multline}
\text{\emph{Stage I}}: \quad \argmin_{\theta, \lambda} \Big(500(A_\lambda x(\theta) - M)^2 +\\
6.81 (\dwelltime \cdot \frameintegration)\Big)
\end{multline}

\begin{multline}
\text{\emph{Stage II}}:\quad \argmax_{\theta^*, \lambda^*} \Big( \textbf{time} (x(\theta),  A_\lambda x(\theta))\Big)\\
\text{report bias, standard deviation of setup.}
\end{multline}

 \noindent was used for the study. Here, $x(\theta)$ is the image taken of the sample, and $\theta$ is the fabricated linear scan top XY AM part [\cite{shao2020effect}]. $M$ is the agreed-upon estimate of what the $\alpha$-lath thickness in the Ti64 sample is, obtained from the workflow used in Shao 2020 [\cite{shao2020effect}]. $A_\lambda$ is the imaging, denoising, and segmentation process being considered, and $A_\lambda x(\theta)$ is the $\alpha$-lath thickness extracted from the image. We place the greatest emphasis on the comparison between $A_\lambda x(\theta)$ and $M$, as evidenced by the weight of 500 on the first term.

 Given a fixed model complexity value, the goal is to minimize the cost to acquire data, and the deviation between collected data and agreed-upon metric $M$. High-resolution images and derived information from Shao et al were used as ground-truth $M$.

 In Table 1 below, one image $x(\theta)$ for each setting $(\theta,\lambda)$ was taken. The $\alpha$-lath thicknesses $A_\lambda x(\theta)$ for each of the dwell times and number of integrated frames (frame integration, or FI) were extracted from the images through an image processing workflow $A_\lambda$. Here, we fix $\lambda$ by using the same image processing workflow for all images. Finally, the $(\alpha)$-lath thickness was extracted from an image at 30 $\mu$s dwell time, 1 FI, and 768 x 512 pixel image resolution to act as the ground-truth setting $M$. The resulting scores were calculated using the objective function defined above.

\begin{table}[hb]
    \centering
    \caption{\emph{Stage I} values for workflows with varying frame integration (FI) and dwell time (DT) settings. Smaller, more negative values indicate higher-quality workflows. We select three workflows with \emph{Stage I} values suitably small and compare them in \emph{Stage II}.}
    \begin{adjustwidth}{-0.3in}{-0.3in}
    \begin{tabular}{@{}p{0.1em}p{2em}ccccccccccc@{}}
    \toprule
     {} & {}& \multicolumn{11}{c}{768 x 512 Pixel Resolution: Dwell Time ($\mu s$) } \\ 
      {}& {}  & 0.3 & 0.4 & 0.5 & 0.6 & 0.7 & 0.8 & 0.9 & 1 & 1.5 & 2 & 3 \\ \cmidrule(){3-13}
        \multirow{7}{*}[-0.3em]{{\rotatebox[origin=c]{90}{Frame Integration}}}& 1 & 4704 & 11296 & 12456 & 12762 & 13673 & 16349 & 12866 & 4649 & 1331 & 803 & 226 \\ 
     &   2 & 4700 & 9171 & 4866 & 3463 & 3204 & 2213 & 3818 & 357 & 636 & 174 & \colorbox{cyan}{42} \\ 
      &  4 & 2247 & 2984 & 834 & 2331 & 720 & 840 & 190 & 95 & 216 & \colorbox{cyan}{64} & 83 \\
      &  8 & 490 & 157 & 174 & 453 & 151 & 148 & 119 & 84 & 100 & 117 & 164 \\ 
      &  16 & 200 & 94 & \colorbox{cyan}{95} & 97 & 95 & 147 & 124 & 114 & 164 & 218 & 330 \\
      &  32 & 87 & 114 & 120 & 154 & 173 & 194 & 226 & 218 & 328 & 437 & 664 \\ 
      &  64 & 146 & 197 & 245 & 274 & 318 & 357 & 401 & 436 & 654 & 872 & 1313 \\ 
      \bottomrule
    \end{tabular}
    \end{adjustwidth}
\end{table}
 Table 1 shows \emph{Stage I} results of the workflow search. As per the framework, the setup corresponding to the lowest objective function value in this figure should not immediately be taken to be the most optimal value. However, based on this \emph{Stage I} search, it is obvious that some setups are more preferable than others. One should conduct a \emph{Stage II} search with setups close to and equal to the lowest \emph{Stage I} score to investigate which setting is truly preferable. The number of potential setups to check will vary based on the application.

\subsection{Stage II Search}
\label{case:stage-ii}
 For \emph{Stage II}, 3 candidate experimental setups were selected as per the \emph{Stage I} results and compared against the ground-truth experimental setup. Using a fixed 768 x 512 pixel image resolution, these were:

\begin{itemize}
    \item \textbf{Previous Study: 30 $\mu$s dwell time, 1 FI at 3072 x 2048 pixel resolution (204.7 s acquisition time)}
    \item \textbf{Ground-Truth: 30 $\mu$s dwell time, 1 FI at 768 x 512 pixel resolution (11.9 s acquisition time)}
    \item Workflow 1: 0.7 $\mu$s dwell time, 16 times FI at 768 x 512 pixel resolution (4.5 s acquisition time)
    \item Workflow 2: 2 $\mu$s dwell time, 4 times FI at 768 x 512 pixel resolution (3.2 s acquisition time)
    \item Workflow 3: 3 $\mu$s dwell time, 2 times FI at 768 x 512 pixel resolution (2.4 s acquisition time)
\end{itemize}

 A comparison of the workflows’ data streams can be seen in Figure 3.
\begin{figure}[H]
\begin{adjustwidth}{-0.4in}{-0.4in}
\centering
\includegraphics[scale=0.65]{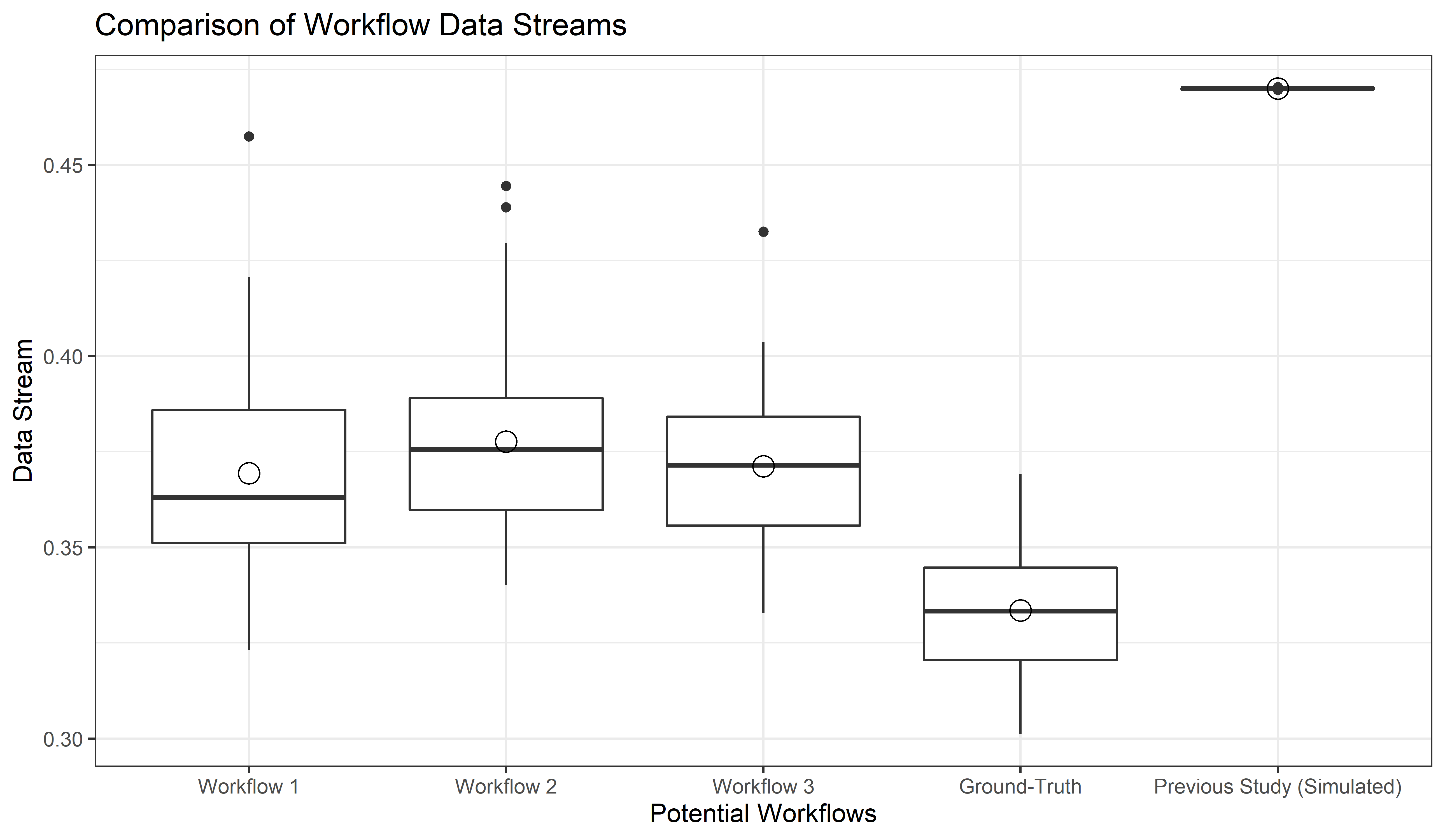}
\caption{Data streams for the candidate workflows as compared to the Ground-Truth workflow. A data stream for the Previous Study workflow was simulated using the statistics reported in \cite{shao2020effect}. The bolded lines represent the median $\alpha$-lath thickness, while the diamonds represent the average $\alpha$-lath thickness. Workflows 1, 2, and 3 all have higher average $\alpha$-lath thickness values than the Ground-Truth Workflow. Workflows 1 and 2 exhibit right skewed distributions, indicating a lower limit in the acceptable signal-to-noise ratio (SNR). The case study's workflows all have a negative bias compared to the Previous Study workflow.}
\label{fig:comp-1}
\end{adjustwidth}
\end{figure}
 We note that all workflows report a higher average $\alpha$-lath thickness than the Ground-Truth workflow, and that Workflows 1-3 \& the Ground-Truth workflow report a lower average $\alpha$-lath thickness what the Previous Study workflow reported for the same sample [\cite{shao2020effect}]. Additionally, Workflows 1-3 also report a higher $\alpha$-lath thickness standard deviation than the Ground-Truth and Previous Study workflows. Lastly, Workflows 1 and 2 exhibit some right-tailed skewness (as their mean is greater than their median), indicating that results achieved using these workflows will be influenced by the SNR lower limit for this system.

 The question is whether or not the information gathered using the three chosen candidate setups will be meaningfully different from one another. After conducting a repeated measures ANOVA (and checking all assumptions-—no extreme outliers, normality in the response, and sphericity of variance), it was concluded that the setups deliver similar results (see Table 2). Therefore, Workflow 3 can safely be chosen as the most valuable high-throughput workflow (Workflow 3: Average $\alpha$-lath thickness = 0.37 \(\mu m\), Standard Deviation = 0.02 \(\mu m\)) (See Table 3 for comparison statistics of Workflow 3's data stream).

\begin{table}[hb]
\centering
\caption{Data streams for the candidate workflows as compared to the ground-truth workflow. The bold lines represent the median $\alpha$-lath thickness, while the diamonds represent the average $\alpha$-lath thickness. Workflows 1, 2, and 3 all have higher average $\alpha$-lath thickness values than the Ground-Truth Workflow. Workflows 1 and 2 exhibit right skewed distributions, indicating an signal-to-noise ratio (SNR) lower limit.}
\begin{tabular}{cc}
\toprule
           Comparison  & P-value \\ \midrule
0.7 $\mu$s - 2.0 $\mu$s & 0.01 \\
0.7 $\mu$s - 3.0 $\mu$s & 1.00 \\
2.0 $\mu$s - 3.0 $\mu$s   & 0.50 \\
\bottomrule
\end{tabular}
\end{table}

\begin{table}[hb]
\centering
\caption{Reported Bias and Standard Deviation differences between the data streams of Workflow 3 and the Ground-Truth workflow, and the data streams of Workflow 3 and the Previous Study workflow.}
\begin{tabular}{ccc}
\toprule
           Workflows Being Compared & Bias \(\mu m\) & Standard Deviation Change \(\mu m\) \\ \midrule
Workflow 3 - Ground-Truth & +0.03 & +0.005 \\
Workflow 3 - Previous Study & -0.10 & +0.01 \\
\bottomrule
\end{tabular}
\end{table}

\section{Discussion}
\label{disc}
 The case study is an example where the cost of acquiring information equivalent to the previous study is very high. As the \emph{Stage I} objective function yields workflows that balance accuracy with cost and complexity, the difference in determined Ti64 $\alpha$-lath thickness between Workflow 3, the Ground-Truth workflow (shown in \emph{Stage II}), and the Previous Study workflow is accounted for. Modifying image processing workflows to match the measurement from the previous study is difficult, as the resolutions between the two studies’ ground-truths are different (previous study: 3072 x 2048, this study: 768 x 512). When this happens, a lower-quality workflow resulting in increased bias or variance of the data stream must be used in order to proceed with the framework. In this case study, we focus on calibrating our potential high-throughput workflows on the specified Ground-Truth workflow for this case study, and simply report the bias and variance of each workflow based on the reported Ground-Truth workflow as well as the Previous Study workflow. As long as data collection using the proposed new workflow proceeds with an understanding of the bias and variance between workflows, continuity and reproducibility between both experiments is achieved.

 The framework also can be generalized to yield high-accuracy workflows and low-variance workflows due to the uncertainty quantification measures and accuracy estimates that the \emph{Stage II} selection process yields. Different objectives will yield different definitions of $\Actionability$, and therefore the result of a \emph{Stage II} workflow will vary based on the user-defined objective. The schematic in Figure 4 shows potential workflows that could be selected based on different objectives. Workflow 1 might be chosen to minimize time, Workflow 2 might be chosen to minimize bias, Workflow 3 might be chosen to balance collection time and bias, and finally the Ground-Truth workflow might be chosen to provide an exact measurement.

\begin{figure}[H]
\centering
\includegraphics[scale=0.65]{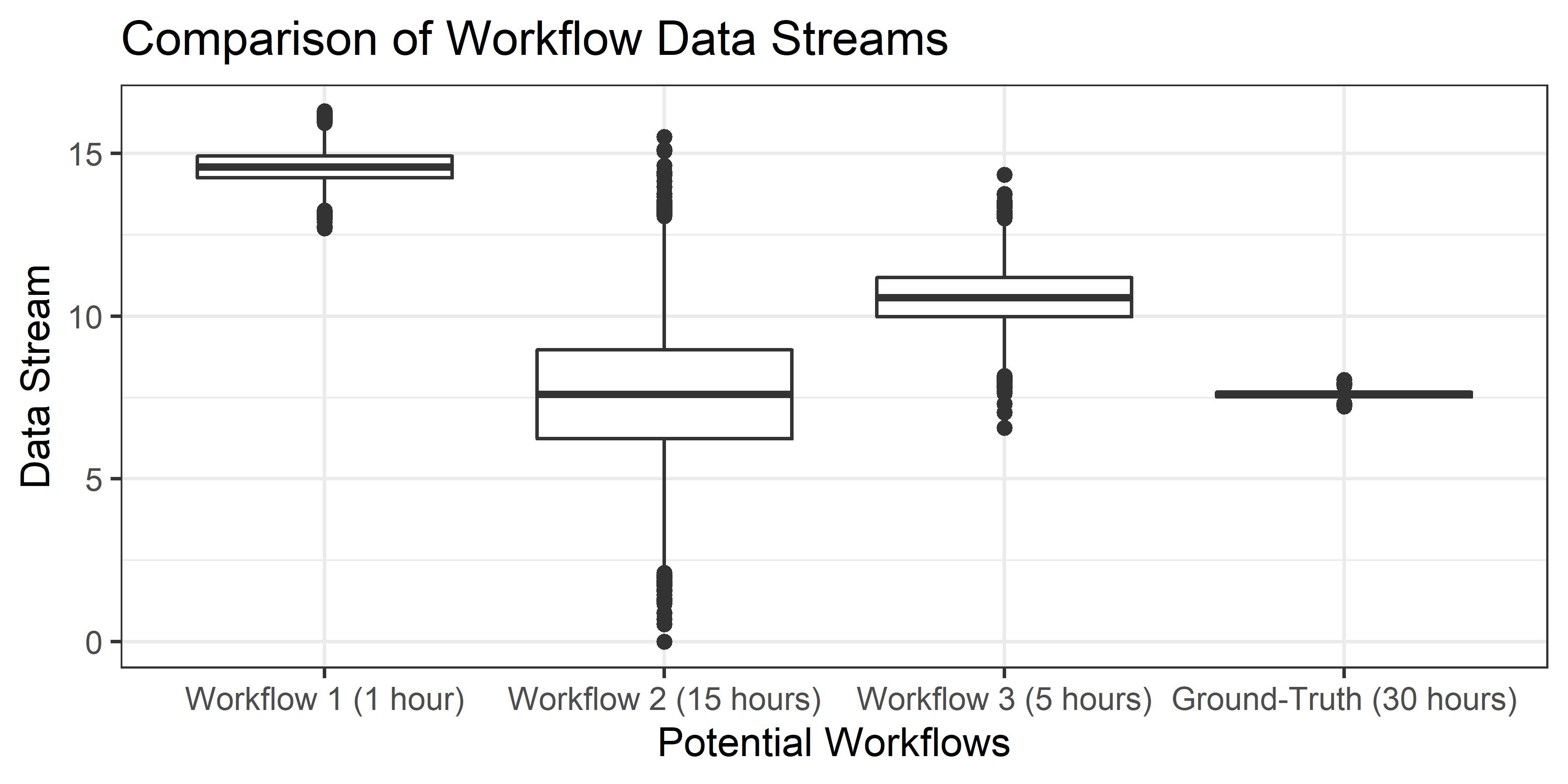}
\caption{Examples of workflow setups that can each be considered optimal given different definitions of actionability. Workflow 1’s variance is low, but the bias is large—this would have to be documented if Workflow 1 is selected for use. Workflow 2’s variance is high but has no bias. This workflow might be considered if accurate results are required at a faster pace than the ground truth. Workflow 3 strikes a balance between the previous two workflows, with a moderate variance and bias. Additionally, it may be decided that exact measurements are required and so the Ground-Truth workflow may also be selected.}
\label{fig:comp-2}
\end{figure}

 The weights for the \emph{Stage I} objective function are to be chosen by human beings, and not the automated system. This is intentional—changing the weights of the parameters of the \emph{Stage I} objective function filters conditions that do not reflect outcomes of that support the posed objective. This allows the framework to yield a customized set of high-value setups with varying accuracy, cost, and complexity. Larger weights on a term will lead to more importance of that term in the overall loss function. For example, assigning larger weights to FI will skew the optimal results to conditions that incorporate lower FI. In practice this choice of weights adds flexibility and allows one to prioritize conditions that best fit their requirements. As an example, investigations of critical components in aerospace applications demand a very accurate estimate of the ground-truth but still want to take advantage of the framework presented. In such a case, we would suggest assigning a very large weight ($>$ 1000) to the bias term, while comparatively smaller terms are assigned to the cost and complexity terms. The \emph{Stage I} objective function will naturally select workflows that collect high-accuracy information while prioritizing the cost to acquire data and complexity to extract information from the data less.

 One of the limitations of the framework is that it does not generalize well to large parameter spaces in the basic case. As an example, testing 10 possible settings each with 10 different possible values yields a \emph{Stage I} search space of $10^{10}$ workflows, which is unacceptably large. This issue can be tackled in three ways. First, by being more selective with the candidate setups in the final step of \emph{Stage I}, one only chooses setups exactly at the minimum or choosing setups with scores having very small differences from the minimum. Second, using prior knowledge, the \emph{Stage I} exploration space can be dramatically reduced to setups that only make sense based on experience. Third, tweaking the weights of the \emph{Stage I} objective function can penalize expensive setups heavily and restrict the exploration space even further. Fourth, the \emph{Stage I} search task can be treated as a multi-armed bandit problem, where an exploratory policy is favored, leading to vast improvements in \emph{Stage I’s} completion speed.

 Lastly, we envision this framework for assisting with the growing problem of data continuity and reproducibility within the scientific community. Historically, results from scientific studies could not be reproduced, either by the authors of the study or by other members in the community. This was due to the necessary differences in the workflows that are employed: methods, capabilities, and tools are subject to change across different research groups and across time. This complicates efforts to establish processing and property standards for materials systems, which is a significant impediment to expediting and optimizing materials development cycles. This framework creates an interface for different workflows to be compared and evaluated—a point of crucial importance as we enter a more mature era of ICME. Using this framework, different workflows can be compared on the \emph{Stage I} objective space as long as they produce the same information for the same objective. Additionally, the standard bias and uncertainty measures that a \emph{Stage II} search provides an easily interpretable set of measures to judge the quality of workflows based on the information they produce.

\section{Conclusion}
\label{conc}
 It has been demonstrated in literature that autonomous experimentation (AE) systems have incredible value to add to the community. For that potential to be realized, we recognize that informatics tools have to become practically implementable for experimenters and a shift in how data is perceived is required. The introduction of new tools requires a means to be evaluated in an experimental setup in order to achieve complete continuity and reproducibility between studies.

 To achieve this, we designed a robust algorithmic framework for the selection of high-throughput workflows that can be completed by AE systems independent of human beings. We used the framework to select the \emph{best user-defined} high-throughput workflow for material characterization on an AM Ti64 sample for the purposes of setting up a basic example. The collection time of BSE-SEM images was reduced by a factor of 5 compared to the Ground-Truth workflow for this case study, and 85 times as compared to the Previous Study workflow. The reported results on the measured Ti64 $\alpha$-lath thickness were still in agreement. 

 Future work will involve utilizing the here presented framework in further studies on metal AM components to develop databases that are critically important to understanding underlying structure-property relationships. Particularly, further focus will be on automating the image pre-processing steps when performing materials characterization experiments to a greater extent.


\bibliographystyle{unsrt}


\end{document}